\documentclass[journal,12pt,onecolumn,draftclsnofoot,]{IEEEtran}

\usepackage{graphicx}
\usepackage{amsmath}
\usepackage{relsize}
\usepackage{booktabs}
\usepackage{subcaption}
\usepackage{amsmath}
\usepackage{algorithm}
\usepackage{xcolor}
\usepackage{blindtext}
\usepackage{enumitem}
\usepackage{amsthm}
\usepackage{amssymb}
\usepackage{stmaryrd}
\usepackage{algpseudocode}
\usepackage{mathtools}
\usepackage{multirow}
\algnewcommand{\IIf}[1]{\State\algorithmicif\ #1\ \algorithmicthen}
\algnewcommand{\EndIIf}{\unskip\ \algorithmicend\ \algorithmicif}

\begin{document}

\title{Predicting Device-to-Device Channels from Cellular Channel Measurements: A Learning Approach}
\author{Mehyar~Najla,~\IEEEmembership{Student Member,~IEEE,}
        Zdenek~Becvar,~\IEEEmembership{Senior Member,~IEEE,}
        Pavel~Mach,~\IEEEmembership{Member,~IEEE,}
        and~David Gesbert,~\IEEEmembership{Fellow Member,~IEEE,}
\thanks{M. Najla, Z. Becvar and P. Mach are with the Faculty of Electrical Engineering, Czech Technical University in Prague, Czech republic
e-mails:  \{najlameh, zdenek.becvar, machp2\}@fel.cvut.cz}
\thanks{D. Gesbert is with the Communication Systems Department, EURECOM, Sophia Antipolis, France
e-mail:  david.gesbert@eurecom.fr}}


\maketitle

\begin{abstract}

Device-to-device (D2D) communication, which enables a direct connection between users while bypassing the cellular channels to base stations (BSs), is a promising way to offload the traffic from conventional cellular networks. In D2D communication, one recurring problem is that, in order to optimally allocate resources across D2D and cellular users, the knowledge of D2D channel gains is needed. However, such knowledge is hard to obtain at reasonable signaling costs. In this paper, we show this problem can be circumvented by tapping into the information provided by the estimation of the cellular channels between the users and surrounding BSs as this estimation is done anyway for a normal operation of the network. While the cellular and D2D channel gains exhibit independent fast fading behavior, we show that average gains of the cellular and D2D channels share a non-explicit correlation structure, which is rooted into the network topology, terrain, and buildings setup. We propose a machine (deep) learning approach capable of predicting the D2D channel gains from seemingly independent cellular channels. Our results show a high degree of convergence between true and predicted D2D channel gains. The predicted gains allow to reach a near-optimal communication capacity in many radio resource management algorithms.
\end{abstract}

\section{Introduction}
In device-to-device (D2D) communication, data is transmitted over a direct link between a pair of nearby user equipment (UEs) instead of being relayed via a base station (BS)\cite{d2d_magazine},\cite{d2d_survey}. Conventionally, the D2D pairs can exploit two communication modes: shared and dedicated \cite{CSD}. In the shared mode, the D2D pairs reuse the same radio resources as cellular users (CUEs) that send data through the BS \cite{SM}. On the contrary, the D2D pairs in the dedicated mode are allocated with resources that are orthogonal to the resources of CUEs \cite{DM}.

An efficient exploitation of the D2D network often entails challenging radio resource management (RRM) problems, such as, selection between shared and dedicated modes \cite{DM}-\cite{Mselection4}, interference management to/from CUEs \cite{Interman1}-\cite{Interman4}, channels and power allocation \cite{CA1}-\cite{binary}, to name just a few. Conventional algorithms addressing the above RRM problems in D2D networks assume a prior estimation of the D2D channel gains (i.e., channel gains among all UEs involved in D2D). In some cases, the full knowledge can be relaxed to a partial knowledge, where only a subset of the distributed D2D channel gains is required (e.g., in \cite{CA6}). Nevertheless, even the partial knowledge of the D2D channel gains implies a substantial cost in terms of an additional signaling overhead on top of the one generated in classical cellular communications. In fact, the cellular channel gains (i.e., channel gains between the UEs and the BSs) are typically estimated by default as these are needed for handover as well as user attachment, authorization, and  classical cellular communication purposes. More precisely, even the users that wish to engage in D2D communications must be recognized by the network and thereby their cellular channel gains must be estimated initially. Thus, these cellular channels are periodically reported to the BSs, and can be leveraged at no additional signaling overhead. An interesting question then arises as to whether the by-default cellular channel gains carry information that is relevant to D2D communication and somehow could help ''for free' to solve the D2D resource management problems.

The idea set forth in this paper is that, while the cellular channel gains should exhibit fading coefficients that are known to be independent of those measured on the direct channels among the UEs, there actually exists common information between these data at the statistical level. In order to build up the reader's intuition, consider the following toy example. Imagine a green-field (free space) propagation scenario, in which the location of all UEs is made available to the network (even for those devices not interested in communicating with the network), then both the cellular and the D2D channel gains would be easily predictable from the UEs' locations and the use of a deterministic free-space channel model with line of sight (LOS) among all entities. Therefor, in a LOS environment, both D2D and cellular channel gains directly relate to each other via the user location knowledge. In practice, however, the UEs' locations may not be known due to privacy issues or may not be simply available. More importantly, in non-line of sight (NLOS) scenarios (such suburban or urban areas), the D2D channels and the cellular channels may be obstructed in completely independent manners making the channel prediction from the UEs' locations seemingly impossible. For instance, two devices might experience a strong LOS D2D channel while a  building may block the cellular channel between one of these devices (or more) and a given BS, thus making the D2D and cellular channel gains seemingly quite a bit less related than in the pure LOS scenario. In this paper, we show that, in contrast to initial belief, a powerful correlation between the cellular and the D2D channels still exists in the NLOS case, and can be made even stronger by leveraging cellular measurements from additional surrounding BS. In this paper, we exploit this correlation through the use of a machine learning approach based on a deep neural network (DNN), which predicts the D2D gains from the cellular gains. Another interesting by-product of our prediction scheme lies in seeing that the set of cellular gains often constitute an order-of-magnitude smaller dimensional object than the D2D channel that we are trying to predict (i.e., there are just X cellular gains for one cell with X users in it, in contrast to X(X-1) direct and interference D2D gains). Hence, the proposed approach not only offers to capitalize on easier-to-get information (cellular channel estimation) rather than on the harder to get D2D channel gains for an optimization of D2D communications, but it also promises substantial savings in signaling for the channel estimation.

In the literature, existing channel prediction works related to this paper typically focus on predicting the channel quality between a single UE and an antenna at the BS at a specific frequency based on either: i) knowing the channel between this UE and the BS antenna at another frequency \cite{Ch_on_ch1}-\cite{Ch_on_ch10}, or ii) knowing the channel between this BS antenna and another UE that is close to the original UE \cite{Ch_on_ch11}, or iii) knowing the channel between this UE and another close-by antenna at the same BS \cite{Ch_on_ch12}. However, the problem presented in this paper, which is predicting D2D channel gains based on the cellular channel gains, is of a different nature from the above-mentioned prediction problems solved in the literature because a strong commonality of space can't be relied upon. Note that this paper builds on and extends our previous work presented in \cite{GLOBECOM}, where we introduced the idea of the DNN-based prediction of the transmission powers for D2D communication. 
Instead, in this paper, we generalize the problem to predicting directly the D2D channel gains. This allows for a more powerful framework, which yields applications to various radio resource management (RRM) related optimization problems in D2D networks.   

The main contributions of this paper are summarized as follows:

\begin{itemize}[leftmargin=0.4cm]
\item We present a novel framework for the D2D channel gains prediction based on the cellular channel gains in order to solve various problems related to radio resource management in D2D communication without incurring the pilot overhead that is usually expected in D2D communication.
\item 	We design a DNN to build up a regression model connecting the cellular channel gains (as DNN inputs) to the D2D channel gains (as DNN outputs). Our results show a high convergence between the true and the predicted D2D channel gains, even in typical urban NLOS scenarios.
\item We demonstrate the efficiency of the proposed framework by applying the predicted D2D gains to existing  channel allocation and power control algorithms presented \cite{MMwcl} and \cite{binary}, respectively. 
\item We analyze the signaling overhead in terms of the number of channel gains needed to implement the radio resource management algorithms from \cite{MMwcl} and \cite{binary} with and without the proposed DNN-based D2D channel gains prediction scheme to show the benefits of the proposed concept. 
\end{itemize} 
	
The rest of the paper is organized as follows. In Section \ref{system}, we present system model and formulate the problem of D2D channel gains prediction. Then, Section \ref{proposed} describes the proposed DNN-based scheme for the prediction of D2D channel gains. Performance evaluation and simulation results are illustrated in Section \ref{performance}. Finally, Section \ref{conclusion} concludes the paper.

\section{System model and problem formulation}\label{system}
In this section, we present our system model, and then, we formulate the problem of the D2D channel gains prediction.

\subsection{System model}\label{sys1}
\begin{figure}[!b]
\centering
  \includegraphics[scale=0.8]{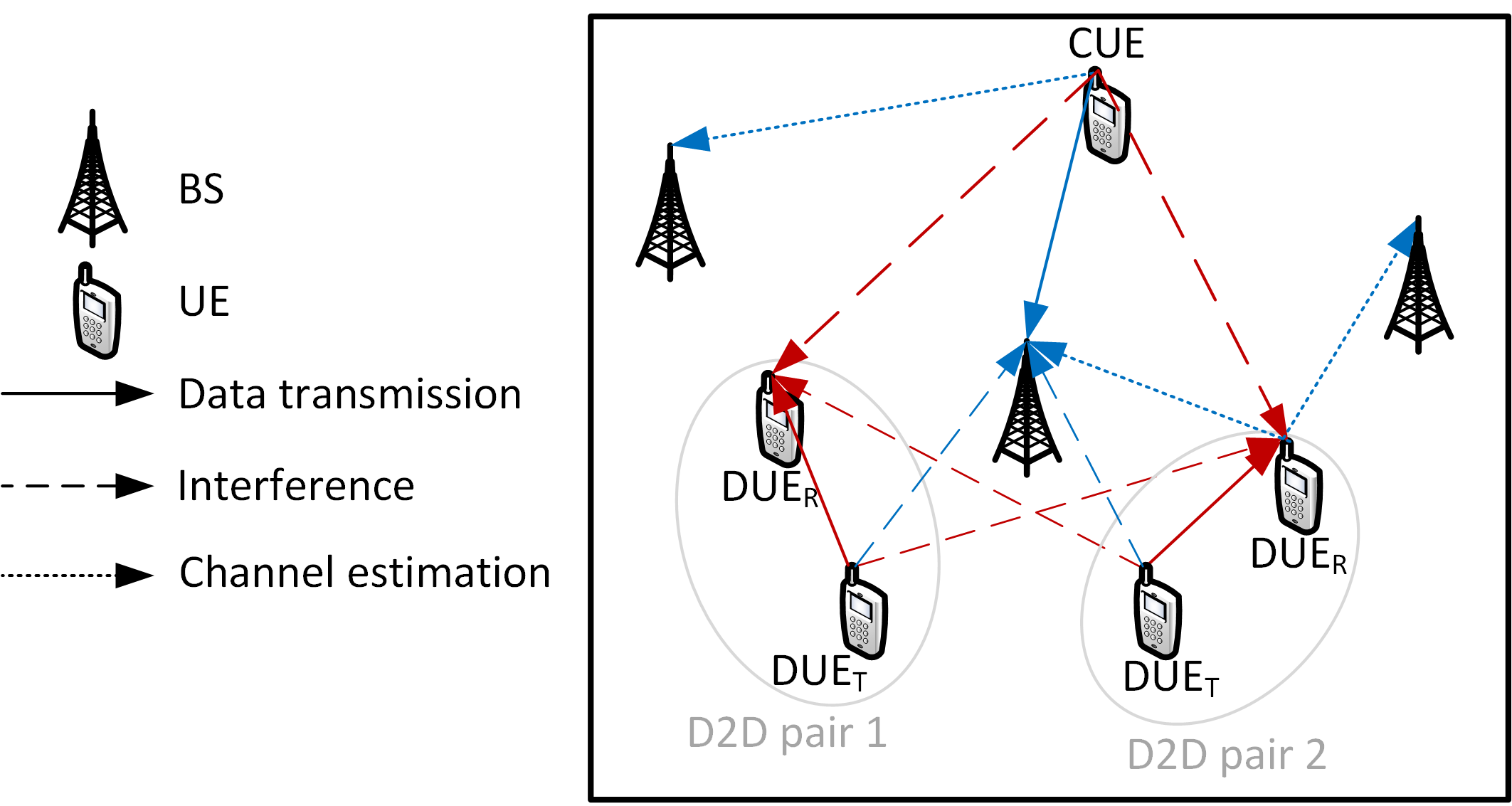}
  \caption{System model: An example with four DUEs, one CUE and three BSs. Note that red and blue colors are used for D2D and cellular channels, respectively, and only part of the signaling (channel estimation) is shown for sake of clarity.}
  \label{system_model_fig}
\end{figure}

In our model, we consider $L$ base stations (BSs) deployed randomly in a square area together with $U$ UEs as shown in Fig. \ref{system_model_fig}. The UEs are divided into $M$ CUEs and $2N$ D2D user equipments (DUEs) composing $N$ D2D pairs, hence, $U=2N+M$. Each D2D pair is composed of a transmitter, DUE$_\text{T}$, and a receiver, DUE$_\text{R}$. 

In general, the capacity of the $n$-th D2D pair at the $k$-th communication channel is defined as:
\begin{equation}\label{cap_calc}
C_{n}^k=B_k log_2\left(1+\frac{p^k_{n} \: g_{n,n}}{B_k\sigma_o+\sum^{q=N}_{\substack{q=1\\q\neq n}}p^k_{q}g_{q,n}+\sum^{m=M}_{m=1}p^k_{m}g_{m,n}}\right)
\end{equation}
where, for the $k$-th channel, $B_k$ is the channel bandwidth, $p_n^k$ is the transmission power of the DUE$_\text{T}$ of the $n$-th D2D pair, $p_m^k$ is the transmission power of the $m$-th CUE, and $p_q^k$ is the transmission power of the DUE$_\text{T}$ of the $q$-th D2D pair causing interference to the $n$-th D2D pair (i.e., $q\in \{1,…,N\}/\{n\}$). Further, $g_{n,n}$ represents the channel gain between the DUE$_\text{T}$ and the DUE$_\text{R}$ of the $n$-th D2D pair, $\sigma_o$ is the noise density, $g_{m,n}$ is the interference channel gain between the $m$-th CUE and the DUE$_\text{R}$ of the $n$-th D2D pair, and  $g_{q,n}$ is the interference channel gain between the DUE$_\text{T}$ of the $q$-th D2D pair and the DUE$_\text{R}$ of the $n$-th D2D pair.

This paper assumes a complete absence of channel gains knowledge among the UEs. Thus, the channel between DUE$_\text{T}$ and DUE$_\text{R}$ of the same D2D pair, interference channels among DUEs of different D2D pairs, and interference channels among the CUEs and the DUEs (i.e., $g_{n,n}$, $g_{q,n}$, and $g_{m,n}$ in (\ref{cap_calc})) are unknown. 

It is important to remember that the DUEs, not to mention the CUEs, need to to estimate uplink/downlink channels to manage efficiently resource allocation and for handover purposes. Thus, although the D2D channel gains are not known by the network, still, the information on the channel quality between each UE (CUE or DUE) and its neighboring BSs are sent periodically to the serving BS in order to update the network information \cite{CSI_feedback_magazine}. The corresponding estimated channel gain between any $i$-th (or $j$-th) UE and the $l$-th BS is denoted as $G_{i,l}$ (or $G_{j,l}$). These cellular channel gains ($G_{i,l}$  and $G_{j,l}$) are assumed to be represented by uplink channel gains estimated (measured) by the BS using the common way from the existing reference signals \cite{reference_signals}. Nevertheless, it is worth to mention that even downlink channel gains can also be used to estimate quality of cellular channels as the downlink gains can be estimated (measured) by the UEs and fed back to the BS.


\subsection{Problem formulation}

We aim to predict the real (true) channel gain $g_{i,j}$ between any $i$-th and $j$-th UEs, that can be, then, exploited for any existing RRM algorithms. Our goal is to minimize the prediction error and we formulate the problem as: 
\begin{equation}\label{main_problem}
 \min_{g^*_{i,j}}(g_{i,j}-g^*_{i,j})^2
\end{equation}
where $g_{i,j}^*$ is the predicted channel gain between the $i$-th and the $j$-th UEs.
To predict the channel gain between any two UEs, we exploit only the available information about each UE, i.e., cellular channel gains. Therefore, in the next section, we propose a novel DNN-based scheme for the prediction of $g_{i,j}$ relying on the knowledge of the cellular channel gains of the $i$-th and the $j$-th UEs.

\section{Prediction scheme}\label{proposed}
This section describes the proposed scheme for prediction of the D2D channel gains among the UEs. First, the principle of the D2D channel gain prediction is illustrated. Then, the architecture of the proposed DNN is described and the training process is clarified. Moreover, we discuss the signaling overhead reduction reached by the proposed prediction scheme.
\subsection{Principle of DNN-based prediction of D2D channel gains exploiting cellular channel gains}\label{initial}
\begin{figure}[!b]
\centering
  \includegraphics[scale=0.7]{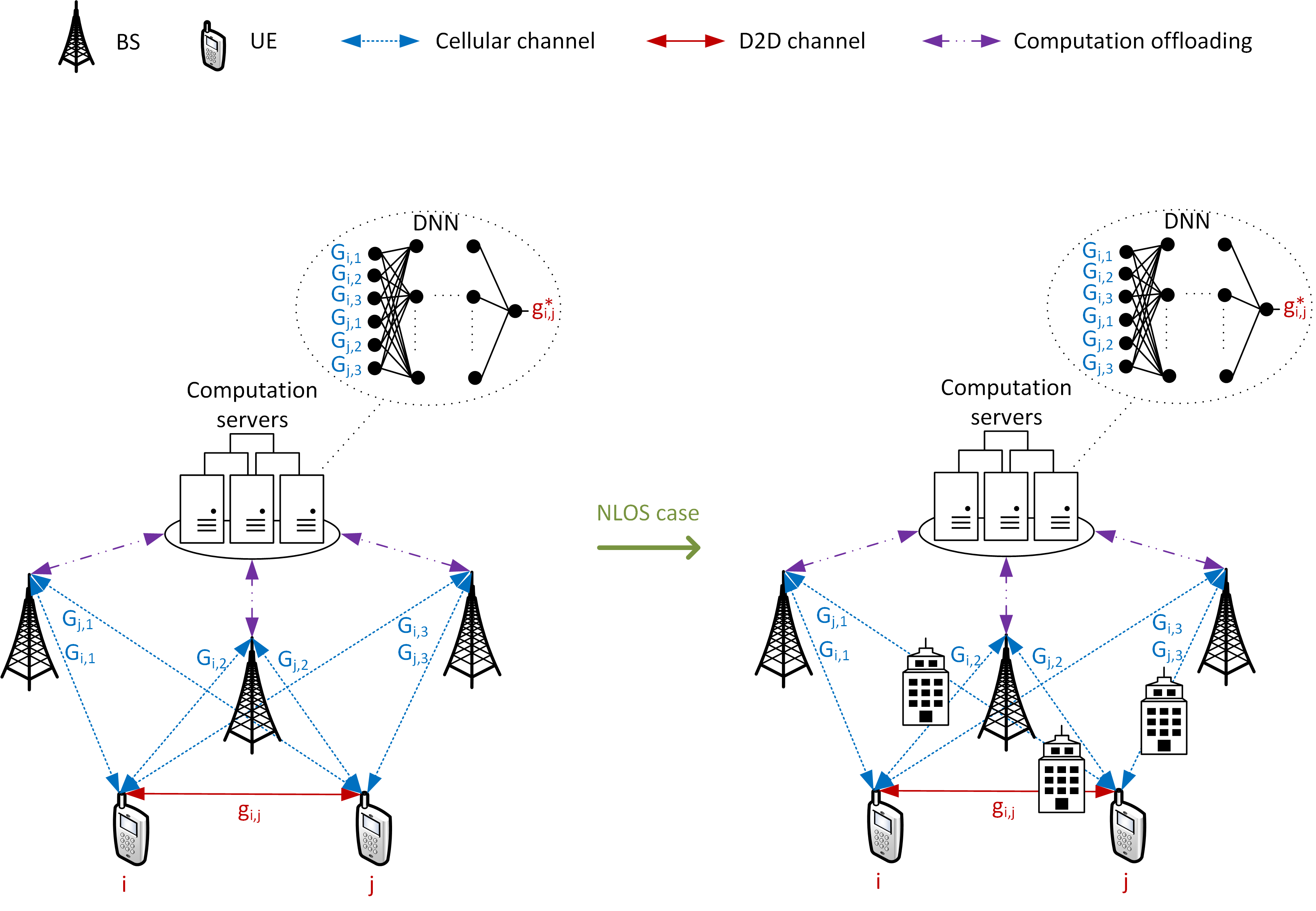}
  \caption{Illustration of D2D channels prediction based on cellular channels for LOS (left part of figure) and NLOS (right part) scenarios}
\label{princip}
\end{figure}

In general, it is clear that in a green-field (free space) propagation scenario, in which the location of all UEs is made available to the network, both the cellular and the D2D channel gains are easily predictable from the UEs' locations. In the free space area with LOS, the cellular channel from the UE to at least three BSs corresponds to a single specific location of the UE. Consequently, the D2D channel gain value between two UEs can be easily predicted in such (unrealistic) scenario. However, in practice, the UEs' locations may not be known due to privacy issues or may not be simply available. Moreover, in NLOS (urban or suburban) scenarios, the D2D channels and the cellular channels may be obstructed in completely independent manner and the D2D channel prediction from the UEs' locations seems to be impossible. For instance, two devices might experience a strong LOS D2D channel while a building(s) obstructs the cellular channel between one of these devices (or more) and the given BS (see Fig. \ref{princip}). In such a case, the D2D channel gains between the two UEs might be hard to predict based on the cellular channel gains. However, in contrast to this initial belief, a powerful correlation between the cellular and the D2D channels is still expected by accounting for additional surrounding BS. The reason behind this is that increasing the number of known cellular channel gains from each UE leads to a higher confidence related to the UE's location and provides information about the position (and shape) of obstructing elements of the terrain. This information can then, in principle, be mapped into a cartography of D2D gains.

To put the above-mentioned intuition into more rigorous terms, given a specific area with certain topology, terrain and buildings' setup, there exists a mapping $\mathbf{F}$ connecting the cellular channel gains of the existing UEs (denoted as $\mathbf{G^C}$) and the D2D channel gains among these UEs (denoted as $\mathbf{g}$) so that:

\begin{equation}\label{mapping_1}
 \mathbf{g}=\mathbf{F}(\mathbf{G^C})
\end{equation}

It is obvious that solving the problem (\ref{main_problem}) can be achieved by approximating the function $\mathbf{F}$ from (\ref{mapping_1}). Nevertheless, this approximation is hard to be done taking into account the changeable size of $\mathbf{G^C}$ and $\mathbf{g}$ when the number of UEs changes. In other words, a different function $\mathbf{F}$ need to be approximated for every possible number of UEs making the solution unrealistic. Therefore, taking into account the problem defined in (\ref{main_problem}), we circumvent this problem by approximating the mapping $F$ between $\mathbf{G^C_{i,j}}$  and $g_{i,j}$ where $\mathbf{G^C_{i,j}}=\{G_{i,1}  ,…G_{i,L}  ,G_{j,1}  ,…G_{j,L}\}$ includes the gains of the cellular channels from $L$ BSs to any $i$-th and $j$-th UEs. In such a way, regardeless of the number of the existing UEs, the D2D channel between any two UEs can be predicted by knowing the gains of the cellular channels from these two UEs and the surrounding BSs. hence, the problem (\ref{main_problem}) is written as:



\begin{equation}\label{problem2}
 \min_{F}(g_{i,j}-F(\mathbf{G^C_{i,j}}))^2
\end{equation}

The optimization problem (\ref{problem2}) aims, by approximating $F$, to minimize the difference between the true (real) and the predicted gains of the D2D channel between any $i$-th UE and $j$-th UE; based on the knowledge of the cellular channel gains of these two UEs.

Deep neural networks are typical up-to-date tools for functions approximation and regression models creation. Thus, in this paper, we exploit the DNN for the prediction of $g_{i,j}$ based on $\mathbf{G^C_{i,j}}$.

Note that, for any UE (DUE or CUE), the cellular channel gains between this particular UE and the surrounding BSs are periodically reported to the BSs for purposes related to the conventional communication and/or handover. In addition, in the future mobile networks, the network computations are supposed to be offloaded to powerful computation servers reducing network’s energy consumption. Those computation servers are located at a relatively far centralized cloud (cloud computing) or placed closer to the UEs, e.g., attached to the BS, in frame of mobile edge computing \cite{MEC}. Thus, the proposed DNN is assumed to take place on the computation servers at the cloud and/or at the edge of the network. The estimated (and reported) cellular channel gains are assumed to be transferred to the computation servers (purple dash-dotted lines in Fig.\ref{princip}), and on these servers, the prediction of $g_{i,j}$ takes place (see Fig. \ref{princip}).


\subsection{The architecture of the proposed DNN}\label{DNN_arch}
\begin{figure}[!t]
\centering
  \includegraphics[scale=0.6]{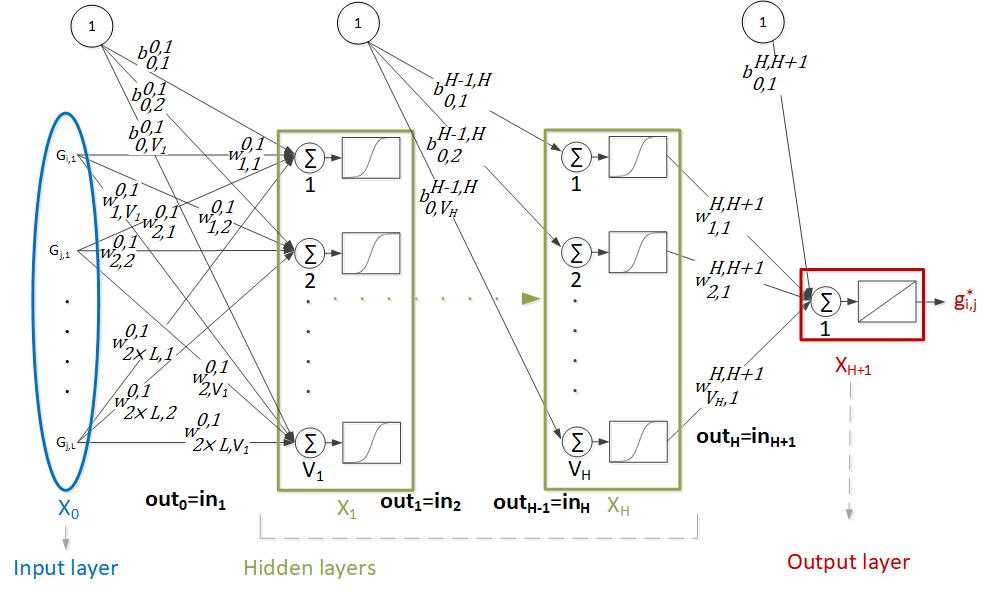}
  \caption{The proposed DNN to build up a regression model connecting input variables (cellular channel gains from two UEs  ($i$ and $j$) to $L$ BSs) and a single output variable (the D2D channel gain between the $i$-th and the $j$-th UE).}
\label{DNN_arch_fig}
\end{figure}

The problem of predicting the D2D channel gain between the $i$-th and the $j$-th UE based on the cellular channel gains from both the $i$-th and $j$-th UEs to the $L$ BSs is a regression problem, which can be solved by the deep neural network designed to build the regression model. Fig. \ref{DNN_arch_fig} shows the proposed fully-connected DNN for regression. The proposed DNN is composed of an input layer ($X_0$), $H$ hidden layers ($X_1, ..., X_H$) and an output layer ($X_{H+1}$). The input layer contains the cellular channel gains between the $i$-th UE and the $L$ BSs and between the $j$-th UE and the $L$ BSs (i.e., $\mathbf{G^C_{i,j}}$) aligned as an input vector in the input layer as illustrated in Fig. \ref{DNN_arch_fig}. Thus, the output of the input layer $\mathbf{out_0}$ is the cellular channel gains vector $\mathbf{G^C_{i,j}}=\{G_{i,1},…G_{i,L}, G_{j,1},…G_{j,L}\}$ of length $2\times L$. Then, the DNN contains $H$ hidden layers whereas every hidden layer $X_h$ is composed of $V_h$ neurons. Every hidden layer $X_h$ has an input vector $in_h$ equivalent to the output of the previous layer $\mathbf{out_{h-1}}$ (i.e., $\mathbf{in_h}=\mathbf{out_{h-1}}, \forall h\in \{1,…,H\}$). Each input element z in $\mathbf{in_h}$ is fed to every neuron $v$ in the hidden layer $X_h$ with a weight $w_{z,v}^{h-1,h}$. Consequently, every neuron $v$ performs dot product between the input elements in $in_h$ and the corresponding weights. The result of the dot product is added to a corresponding bias $b_{0,v}^{h-1,h}$ and processed by commonly used sigmoid function giving the output of the neuron. Hence, the hidden layer $X_h$ with $V_h$ neurons and input vector $\mathbf{in_h}$ gives an output vector $\mathbf{out_h}$ of the length $V_h$ and this output vector $\mathbf{out_h}$  is, thus, written as: 
\begin{equation}\label{out_h}
  \begin{multlined}
\mathbf{out_h}=Sig(\mathbf{W^{h-1,h}}\mathbf{in_h}+\mathbf{b^{h-1,h}})
=Sig(\mathbf{W^{h-1,h}}\mathbf{out_{h-1}}+\mathbf{b^{h-1,h}})
  \end{multlined}
\end{equation}
where $Sig$ is the sigmoid function $Sig(Z)=\frac{1}{1+exp(-Z)}$ , $\mathbf{W^{h-1,h}}$ is the matrix of weights of the links between every input element of $X_h$ (i.e., equivalent to the output of $X_{h-1}$) and every neuron in $X_h$ and $\mathbf{b^{h-1,h}}$ is the vector of biases attached to every neuron. 

The output of the last hidden layer $\mathbf{out_H}$ is followed by the output layer. The output layer in the DNN for regression of a single variable is composed of one neuron. The single neuron of the output layer performs the dot product between $\mathbf{out_H}$ and corresponding weights $\mathbf{W^{H,H+1}}$ (i.e., the vector of weights dedicated to the links between outputs of the last hidden layer $X_H$ and the single neuron in the output layer $X_{H+1}$). Then, the output layer neuron also sums its attached bias scalar $b^{H,H+1}$ and implements a linear activation function giving an output as:
\begin{equation}\label{out_DNN}
  \begin{multlined}
g^*_{i,j}=Lin(\mathbf{W^{H,H+1}}\mathbf{out_H}+{b^{H,H+1}})
  \end{multlined}
\end{equation}
where $Lin$ is the linear activation function $Lin(Z)=Z$ and the output $g_{i,j}^*$ of the proposed DNN is the predicted D2D channel gain between the $i$-th and the $j$-th UEs.

\subsection{Offline learning and exploitation of the proposed DNN  }\label{learning}
We propose a supervised learning-based solution to predict the D2D channel gains based on the cellular channel gains. To this end, the D2D channel gain $g_{i,j}$ between the $i$-th UE and the $j$-th UE is derived offline. Then, $g_{i,j}$ is fed to the proposed DNN as a target attached to the cellular channel gains between the $i$-th UE and $L$ BSs and between the $j$-th UE and $L$ BSs ($\mathbf{G^C_{i,j}}$) presenting features. The features (i.e., cellular channel gains) and the target (i.e., D2D channel gain $g_{i,j}$) compose together a single learning sample. As a pre-training step, learning samples are collected and, then, split into a training set and a test set. The samples from the training set are used to train the proposed DNN while the samples in the test set are used to test the accuracy of the trained DNN on a set of samples that is not used for training to prevent overfitting \cite{overfitting}.
During the training process, a loss function is defined to evaluate the regression model prediction accuracy. The loss function in the DNN that builds the regression model predicting a single variable is, typically, a measurement showing how far is the predicted value of the variable from the true value of this variable ($g_{i,j}^*$ and $g_{i,j}$ in our case). Therefore, taking the optimization problem (\ref{problem2}) into account, we consider a mean square error loss function that can be written as:

\begin{equation}\label{loss_func}
\iota=\frac{1}{S} \sum_{s=1}^{s=S}(g^s_{i,j}-{g^{s*}_{i,j}})
\end{equation}

where $S$ is the number of the training samples, $g_{i,j}^s$ is the target (true D2D channel gain) of the $s$-th training sample, and $g_{i,j}^{s *}$ is the predicted D2D channel gain based on the cellular channel gains of the $s$-th training sample. 

To minimize the mean square error loss function, the weights and biases of the proposed DNN are updated using Levenberg-Marquardt Backpropagation algorithm, which is an optimization method designed to solve non-linear least squares problems \cite{LM1}. Thus, Levenberg-Marquardt algorithm can be applied with backpropagation for the neural networks training when the loss function is a sum of squares \cite{LM2}.

It is worth to mention that the whole learning phase (i.e., collecting samples, training, and testing the proposed DNN) is done offline, i.e., before it is applied to the real mobile network (or before its testing by simulating the mobile network). Therefore, the cellular channel gains derived from the simulations can be used for the offline training and testing and, then, the trained DNN is applied directly in the real network. 

The proposed DNN is able to predict the channel gain between any pair of UEs. Thus, for multiple UEs, the trained (and tested) DNN is utilized to predict all needed channel gains among every pair of UEs independently and in parallel. To be more specific, based on the cellular channel gains of the UEs, we utilize the trained DNN to obtain all D2D channel gains, such as the channel gains between every two DUEs of the same D2D pair, interference channel gains between every couple of DUEs from different D2D pairs and interference channels between the CUEs and the DUEs. These can be, then, exploited to solve any RRM problem using the existing algorithms.

\subsection{Analysis of reduction in signaling overhead}\label{signaling}
In this subsection, we discuss the signaling overhead in terms of the number of channel gains that need to be estimated (measured) in the network.

In the existing network, the cellular channel gains between the UEs and the neighboring BSs are commonly estimated  (i.e., for conventional communication and handover purposes). The number of the commonly estimated cellular gains is $L(2N+M)$. Note that even the DUEs might need to change from the D2D communication to the conventional communication in the case of a sudden D2D communication quality drop and, therefore, the cellular channels of DUEs are also periodically estimated and reported. 

In the literature, for conventional RRM algorithms related to the D2D communication (e.g., power control algorithm from \cite{binary}), additional $2N(2N-1)$ direct and interference D2D channels need to be estimated between the $2N$ DUEs. Moreover, for the D2D in shared mode, interference channels between the CUEs and the DUEs have to be estimated and reported as well. The number of those interference channels between the $M$ CUEs and the $2N$ DUES that should be estimated is $2NM$. Thus, the number of estimated channel gains in the common network with the D2D communication is: 
\begin{equation}\label{Numb_gains_1}
\Sigma = L(2N+M)+2N(2N-1)+2NM
\end{equation}

In this paper, we predict the D2D channel gains from the common estimated cellular gains. In other words, in the network with D2D communication utilizing the proposed prediction scheme, the number of channel gains need to be estimated (measured) is limited to the estimation of $L(2N+N)$ channel gains, which are used to predict the remaining needed D2D channel gains. Thus, by subtracting $L(2N+M)$ from (\ref{Numb_gains_1}), we can calculate the reduction in the number of  estimated channel gains. This reduction, in the shared mode, is equal to:
\begin{equation}\label{Numb_gains_2}
\Delta \Sigma =\Sigma - L(2N+M)=2N(2N-1)+2NM
\end{equation}
In the dedicated mode, the CUEs do not affect the D2D communication as the channels allocated to the CUEs are orthogonal to those allocated to the D2D pairs. In such case, the reduction in the number of estimated channel gains achieved by the proposed prediction scheme is determined by setting $M$ to zero in (\ref{Numb_gains_1}) and (\ref{Numb_gains_2}), respectively.

\section{Performance evaluation}\label{performance}
In this section, we describe the simulation scenarios and parameters, and then, we discuss simulation results from two different perspectives as follows. First, we analyze the accuracy of the prediction scheme statistically showing how close the predicted D2D channel gains are to the true gains of the D2D channels. Second, we illustrate the performance of the proposed prediction scheme on selected examples of existing algorithms for D2D RRM in the mobile network, and we show how this prediction scheme affects the D2D communication quality and network’s signaling overhead. To this end, it is important to remember that the proposed prediction scheme, in this paper, aims to reduce the signaling overhead needed for D2D communication without significant losses in the communication quality.

\begin{figure}[!b]
\centering
  \includegraphics[scale=0.8]{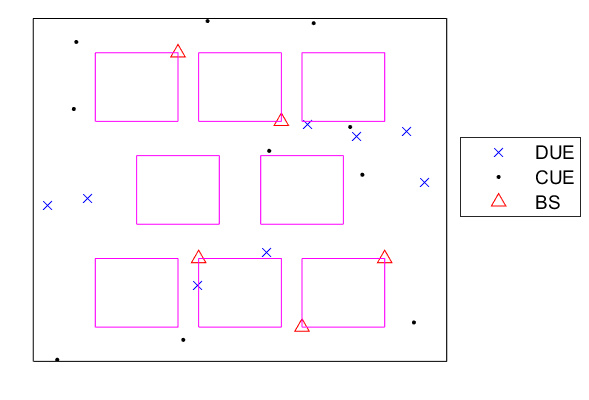}
  \caption{Example of simulation deployment with buildings (pink rectangles) for urban area with $N=4$, $M=10$ and $L=5$. Note that no buildings are present in rural area.}
\label{scenario}
\end{figure}

\subsection{Simulation scenarios and performance metrics}\label{scenarios_subsec}
We consider up to $20$ DUEs (composing up to $10$ D2D pairs) and $10$ CUEs deployed uniformly within an area of $250\times250$ m$^2$ covered by up to $5$ BSs. Although the DUEs are uniformly distributed, the maximum distance between the DUE$_\text{T}$ and the DUE$_\text{R}$ of the same D2D pair is upper-bounded by a maximal distance of $d_{max}=50$ m as in \cite{dmax1}-\cite{dmax2} to guarantee availability of D2D communication. For any D2D transmitter, the maximal and the minimal transmission powers are set to $p_{max} = 24$ dBm and  $p_{min} = 1$ dBm, respectively, like in \cite{GLOBECOM}. 

\begin{table}[t] 
	\centering
	\caption{Simulation parameters.}\label{table1}
	\label{tab:table1}
\begin{tabular}{ |ll|l|}
\hline
\multicolumn{2}{ |c| }{ Parameter}  & {Value}\\ \hline 
Carrier frequency & $f_{c}$ & $ 2 $  GHz \\
Bandwidth & $B$ & $ 20$  MHz  \\
Number of D2D pairs  & $N$ & $ 2-10$ $ $ \\
Number of CUEs (shared mode only)  & $M$ & $ 10$ $ $ \\
Number of channels (shared mode only)&$K$&$10$\\
Bandwidth per any $k$-th  channel (shared mode only)& $B_k$ & $ 2$  MHz  \\
Maximal distance between DUE$_\text{T}$ and DUE$_\text{R}$ of the same pair  & $d_{max}$ & $ 50$~m $ $ \\
Number of BSs  & $L$ & $ 1-5$  $ $ \\
Maximal transmission power  & $p_{max}$  & $ 24$  dBm \cite{GLOBECOM} \\
Minimal transmission power  & $p_{min}$  & $ 1$  dBm \cite{GLOBECOM}  \\
Noise power spectral density&$\sigma_o$ & $  -174$  dBm/Hz \\
\hline
\end{tabular}
\label{Tab01}
\end{table}

We consider two different scenarios according to the signal propagation between the UEs and the BSs and among all UEs. The first scenario assumes an open rural area with full availability of line-of-sight (LOS) for all channels (D2D channels and cellular channels). The second scenario, shown in Fig. \ref{scenario}, presents an urban area (such as scenario C2 in \cite{winner}) with building blocks forming a Manhattan-like grid (see the pink rectangular building blocks in Fig. \ref{scenario}). In the second scenario, the buildings lead to a certain probability of non-line-of-sight (NLOS) for both the D2D and cellular channels. In both rural and urban areas, the LOS path loss is generated in line with 3GPP recommendations \cite{3gpp_D2D_Pl}. In the urban scenario, we assume that the communication channel intercepted by a single or more building walls is exposed to an additional loss of $10$ dB per wall as in \cite{GLOBECOM}. Note that Fig. \ref{scenario} presents a 2D projection of the simulated urban area, nevertheless, in our simulations, the building heights are distributed uniformly between $20$ and $30$ m to randomly affect NLOS and LOS probabilities. Simulation parameters are summarized in Table \ref{Tab01}.

For the learning process, we generate $1~000~000$ learning samples divided into $700~000$ samples used to train the DNN (i.e., the training set) and another $300~000$ samples for testing (i.e., the test set). Note that we show also an impact of the number of learning samples on the prediction accuracy in the next subsections. The proposed DNN exploits five hidden layers composed of $20$, $18$, $15$, $12$, and $8$ neurons, respectively. Note that the number of hidden layers and the number of neurons in each layer are set by trial and error approach.

In this paper, we evaluate the proposed prediction scheme from two following perspectives:
\begin{description}
  \item[$i)$] Statistical evaluations related to prediction accuracy (before implementing the prediction scheme in the mobile network). For the statistical evaluation, we consider the well-known Pearson correlation coefficient as a performance metric to show how close are the predicted D2D channel gains to the true channel gains statistically. It is important to remember that Pearson correlation coefficient values range between zero and one where the value of one represents a complete matching between the predicted and the true values of the D2D channel gains.
  \item[$ii)$] Evaluations related to the D2D communication performance represented by the sum capacity of the D2D pairs $C=\sum_{n=1}^{n=N}\sum_{k=1}^{k=K}C_n^k$. Moreover, we analyze the signaling overhead corresponding to the number of channel gains that need to be estimated/reported in the network.

Both above-mentioned evaluation perspectives are presented in the next two subsections.
\end{description}

\subsection{Statistical analysis of the prediction scheme}

\begin{figure}[!b]
\centering
  \includegraphics[scale=0.8]{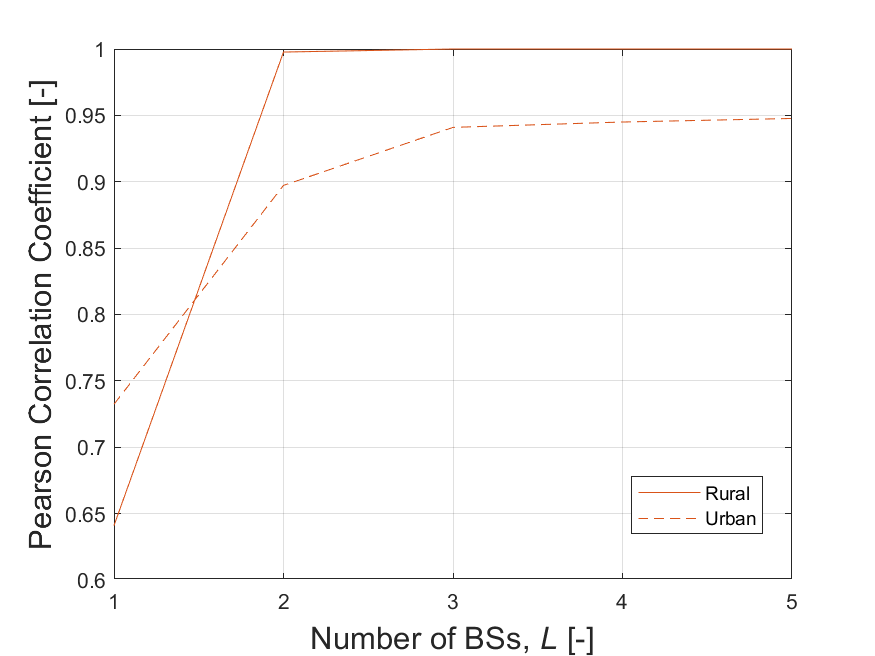}
  \caption{Pearson correlation coefficient between the true and the predicted D2D channel gains as a function of number of BSs $L$}
\label{res_1}
\end{figure}

In this subsection, we analyze the results related to $g_{i,j}$ prediction statistically. In other words, as the training is done offline before its usage in the mobile network, we aim to study the prediction accuracy from the statistical point of view showing how close we expect the predicted gain of a D2D channel to be compared to the true gain of this channel. We show the statistical results of predicting a single D2D channel gain by testing the trained DNN on the test set.

\begin{figure}[b!]
\centering
\begin{subfigure}[t]{0.45\textwidth}
\centering
\includegraphics[scale=0.6]{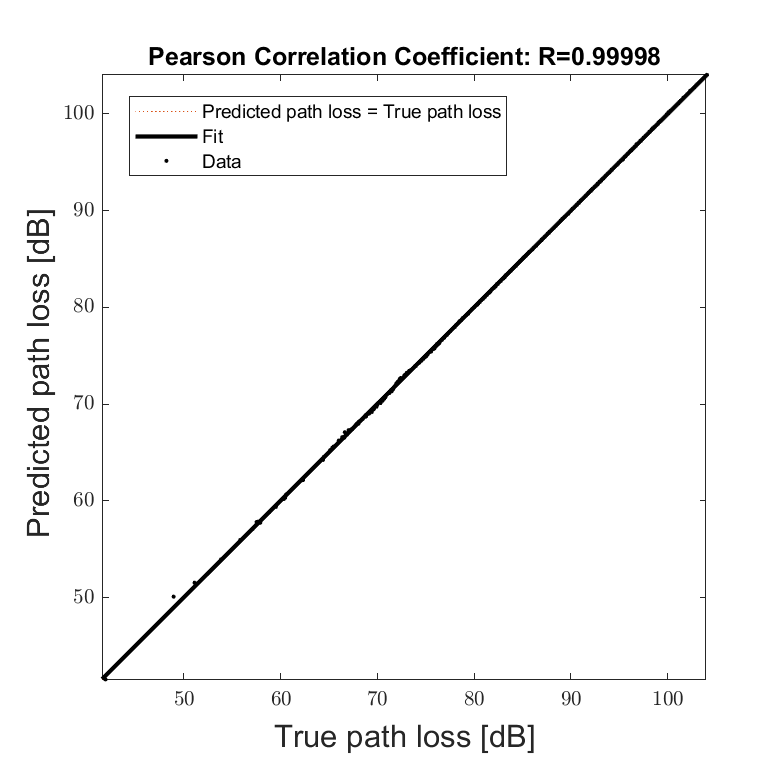}
\caption{ }\label{res_2a}
\end{subfigure}
~~~~~~~
\begin{subfigure}[t]{0.45\textwidth}
\centering
\includegraphics[scale=0.6]{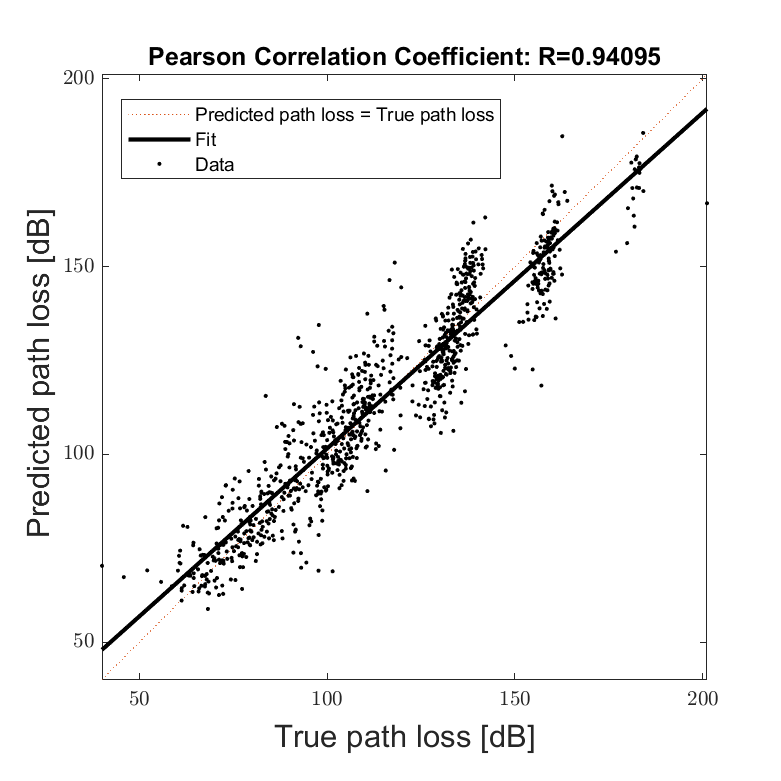}
\caption{ }\label{res_2b}
\end{subfigure}
\caption{Regression plot for rural (a) and urban (b) scenarios for $L=3$ BSs.}
\label{res_2}
\end{figure}

Fig. \ref{res_1} shows Pearson correlation coefficient between true and predicted D2D channel gains over different number of BSs. As expected, for both rural and urban scenarios, the Pearson correlation coefficient increases with the number of BSs. The reason of this increase is that the more BSs in the area, the more information about each UE are known (i.e., we know cellular gains to more BSs). When number of BSs reaches three, the Pearson correlation coefficient saturates to its maximum values of $0.999$ and $0.94$ for the rural and urban scenarios, respectively. The higher value of the Pearson correlation coefficient is reached in the rural scenario because the cellular channel gains are less random in the rural scenario due to absence of the building, and presence of LOS channels only. 

\begin{figure}[!t]
\centering
  \includegraphics[scale=0.8]{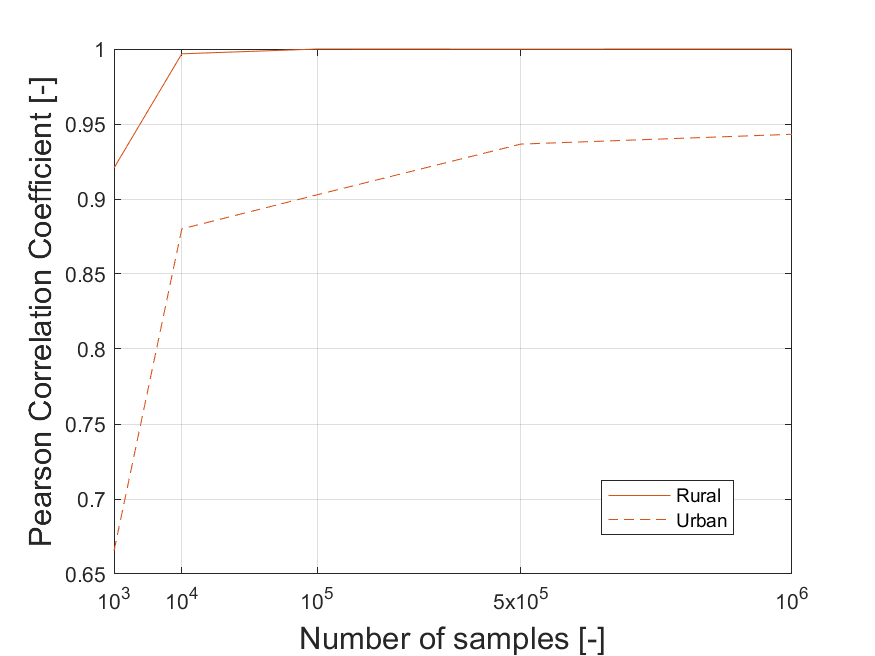}
  \caption{Pearson correlation coefficient between the true and the predicted D2D channel gains as a function of number of learning samples for $L=3$ BSs.}
\label{res_3}
\end{figure}

Fig. \ref{res_2} shows the regression plot for rural (Fig. \ref{res_2a}) and urban (Fig. \ref{res_2b}) scenarios with $L=3$ BSs and considering $1~000$ testing samples from the test set. In general, we can see that the values of the path loss in the urban scenario are spread in a wider domain, because of a presence of the buildings and NLOS as explained in Section \ref{scenarios_subsec}. We can also see, in Fig. \ref{res_2a}, that the predicted path loss (i.e., $10log_{10} (1/g_{i,j}^*)$) matches almost perfectly the true path loss (i.e., $10log_{10} (1/g_{i,})$) for the rural area. In contrast, some deviation of the predicted path losses from the true values can be seen in Fig. \ref{res_2b} in the urban area. This deviation is a result of the existence of the buildings producing some randomness and uncertainty in the values of the estimated channel gains. Nevertheless, the predicted and the true path losses are, still, highly correlated and Pearson correlation coefficient equals $0.94$ even for the urban scenario. 

Note that results presented in Fig. \ref{res_1} and Fig. \ref{res_2} are based on learning with $1~000~000$ samples. Consequently, to illustrate the influence of the number of samples on the learning accuracy, Fig. \ref{res_3} shows Pearson correlation coefficient over number of samples for both rural and urban scenarios. In the rural and urban areas, the correlation coefficient increases with the number of samples rapidly at the beginning for lower numbers of the samples. Then, the correlation coefficient increment with the number of learning samples becomes negligible. We can further see that, in the rural scenario, $10~000$ samples are sufficient to reach almost a perfect matching between the predicted and the true D2D channel gain. However, for the urban scenario, $500~000$ learning samples are needed to reach the correlation coefficient of $0.94$. This can be explained by the higher difficulty of constructing the regression model that connects cellular channel gains to the D2D channel gains in the case where buildings (or obstacles) exist and randomize the path loss values.  

\begin{figure}[!t]
\centering
  \includegraphics[scale=0.8]{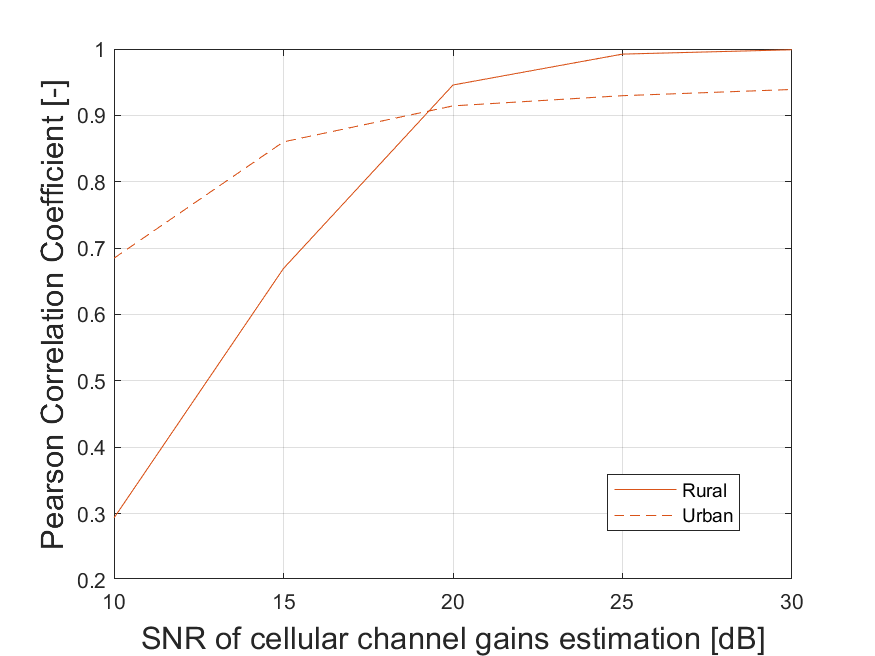}
  \caption{Pearson correlation coefficient between the true and the predicted D2D channel gains as a function of the cellular channel estimation accuracy represented via estimation SNR for $L=3$ BSs.}
\label{res_4}
\end{figure}

In Fig. \ref{res_4}, we show the effect of the possible noise and inaccuracy in the estimation (measurement) of the conventional cellular channels by the BSs. To this end, we define $SNR_G$ as zero-mean Gaussian noise (i.e., the error) added to the modeled cellular channel gain estimation. Hence, $SNR_G$ represents the cellular channel gain estimation accuracy and it is expressed as the ratio between the true cellular channel gain (UE to BS) and the noise representing an error in estimation of the UE to BS channel. Thus, we add the noise of $\mathcal{N}(0,e)$ (where $SNR_G=10log_{10}(\frac{G_{i,l}}{e})$ dB) to the estimated cellular channel gain $G_{i,l}$. As we see in Fig. \ref{res_4}, the correlation coefficient between the true and predicted D2D channel gains increases with the increasing accuracy of the estimated cellular channel gains. If the cellular channels are estimated with $SNR_G$ of $30$ dB, the Pearson correlation coefficient is $0.999$ and $0.94$ for the rural and urban scenarios, respectively. 

\subsection{Performance of D2D communication aided by the prediction scheme}

\begin{figure}[!b]
\centering
  \includegraphics[scale=0.8]{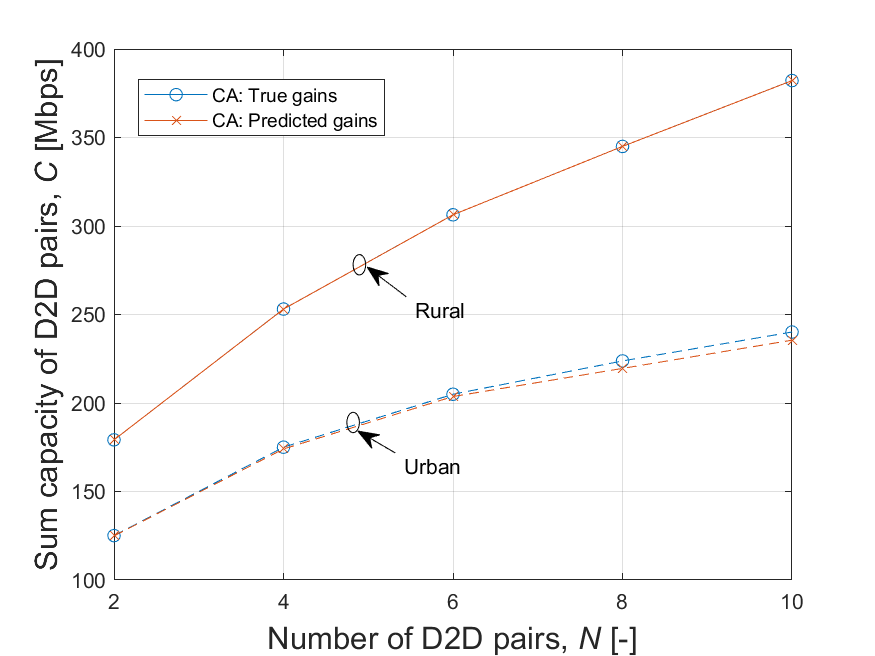}
  \caption{Sum capacity of D2D pairs as a function of number of D2D pairs when channel allocation scheme from \cite{MMwcl} is implemented on true and predicted D2D channel gains with $L=3$ BS and $M=10$ CUEs.}
\label{res_5}
\end{figure}

In this subsection, we show the impact of exploiting the proposed D2D channel prediction scheme based on the machine learning for the D2D communication in the mobile network. For this purpose, we adopt two up-to-date RRM algorithms, one for the channel allocation in the D2D shared mode \cite{MMwcl} and one greedy algorithm for a binary power control in the D2D dedicated mode \cite{binary}. For both algorithms, we compare the performance (i.e., sum capacity of D2D pairs and the number of channels need to be estimated) in the case when these algorithms are supported by our proposed D2D channel prediction scheme with the case when these algorithms are implemented without the machine learning-based prediction approach according to the respective original papers \cite{MMwcl} and \cite{binary}. The purpose of this comparison is to show that the performance of the existing RRM schemes reached with the proposed prediction scheme is not impaired while a substantial reduction in signaling overhead is achieved. Note that, in the legend of this subsection's figures, CA and PC are used to denote channel allocation scheme from \cite{MMwcl} in the shared mode and binary power control from \cite{binary} in the dedicated mode.

 Fig. \ref{res_5} shows the sum capacity of D2D pairs over the number of D2D pairs communicating in the shared mode and with the channel allocation scheme from \cite{MMwcl} implemented on true and predicted D2D channel gains. It is clear that, in the rural area, the predicted channel gains can lead to a sum capacity that matches the one that can be reached if the true channel gains are known for different numbers of D2D pairs (solid lines in Fig. \ref{res_5}). The reason for this outstanding performance in the rural area is the high accuracy of the D2D channel gains prediction as shown in previous subsection. Even in the urban area, we can see only a negligible  loss in terms of the sum capacity (up to $1.9\%$ loss for $10$ pairs) comparing to the channel allocation scheme with full knowledge of the true values of all D2D channel gains. Note that, In Fig. \ref{res_5}, the changes of the sum capacity of D2D pairs over different numbers of D2D pairs, in both scenarios, follows the behavior described in \cite{MMwcl}. 
 
 \begin{figure}[!t]
\centering
  \includegraphics[scale=0.8]{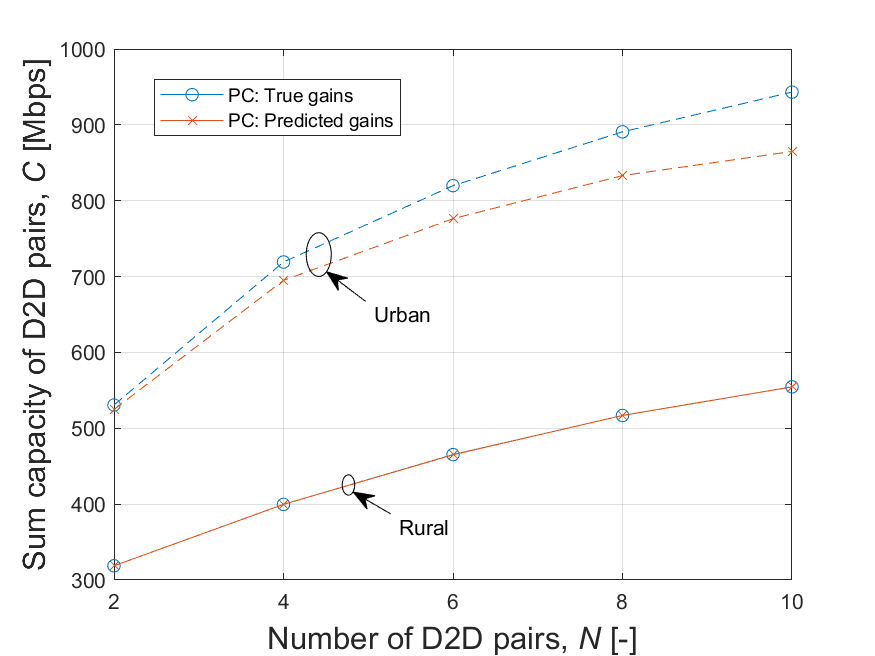}
  \caption{Sum capacity of D2D pairs as a function of number of D2D pairs when greedy algorithm for binary power control from \cite{binary} is utilized to allow all D2D pairs to reuse the whole bandwidth in the D2D dedicated mode ($L=3$ BS).}
\label{res_6}
\end{figure}

The performance of the greedy algorithm for binary power control in D2D dedicated mode from \cite{binary} is shown in Fig. \ref{res_6}, where the D2D pairs are considered to reuse the whole bandwidth. Then, the greedy algorithm is implemented to make a binary transmission power decision for each D2D pair with true and predicted D2D channel gains. In the rural area, a perfect matching between the binary power control implemented on true and on predicted gains is achieved due to the very high accuracy in the prediction of the D2D channel gains. In the urban area, only a small loss in the sum capacity, ranging from $1\%$ (for two pairs) to $9\%$ (for ten pairs), is introduced by implementing the binary power control on the predicted channel gains comparing to the binary power control based on true gains. However, such a loss can be expected by the fact that making a binary decision about the transmission power of each D2D pair is critical and highly sensitive to the accuracy of the predicted D2D channel gains. Nevertheless, ten D2D pairs in proximity reusing a single channel is an extreme case that is not expected to occur often in the real network. In contrast, a reasonable case is when, approximately, four or six D2D pairs reuse a single channel. For four and six D2D pairs, the binary power control implemented on the predicted channel gains loses only $3.4\%$ and $5.6\%$, respectively, comparing to the binary power control implemented on true D2D channel gains.

\begin{figure}[!t]
\centering
  \includegraphics[scale=0.8]{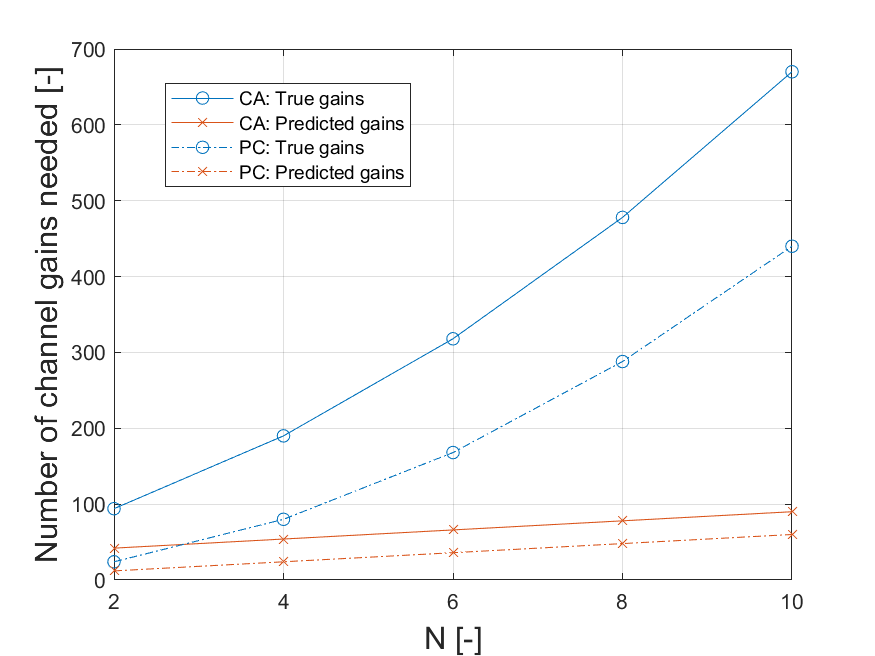}
  \caption{Signaling overhead in terms of number of channels need to be estimated by the network as a function of number of D2D pairs; when channel allocation scheme from \cite{MMwcl} or binary power control from \cite{binary} is implemented on true and predicted D2D channel gains with $L=3$ BS and $M=10$ CUEs.}
\label{res_7}
\end{figure}

In Fig. \ref{res_7}, we show the signaling overhead in terms of the number of channels estimated by the network if the channel allocation scheme from \cite{MMwcl} and the greedy algorithm for binary power control from \cite{binary} are implemented on true and predicted D2D channel gains. As shown in Fig. \ref{res_7}, for both the channel allocation scheme from \cite{MMwcl} and the power control algorithm from \cite{binary}, the number of estimated channel gains with the proposed prediction scheme is significantly lower than when all the channel gains would need to be estimated. More specifically, we need to estimate/report up to approximately seven times less channel gains if the proposed DNN-based prediction is used for the channel allocation scheme from \cite{MMwcl} or the power control algorithm from \cite{binary} comparing to the case when the knowledge of all gains would be required. 

\section{Conclusion}\label{conclusion}
In this paper, we have proposed a novel D2D channel gains prediction scheme based on the cellular channel gains between the UEs and multiple BSs. The proposed prediction scheme takes the advantage of the network topology-related correlation between the cellular and D2D channel gains. Supervised learning-based approach exploiting deep neural networks has been implemented to extract the mapping between the cellular channel gains of any couple of the UEs (i.e., gains of channels between these two UEs and multiple BSs) and the gain of the D2D channel between these two UEs. The proposed prediction scheme achieves a high Pearson correlation coefficient between the true and the predicted D2D channel gains. In addition, we show that the proposed prediction scheme significantly reduces the networks’ signaling (represented by channel state information) overhead if applied to realistic radio resource management algorithms. This saving of the channel information is at the cost of only a negligible performance losses in terms of communication capacity comparing to the conventional implementation of these algorithms with knowledge of all channels.

The future work should focus on improving the prediction scheme performance (prediction accuracy) for scenarios with buildings and obstacles existence. A dynamic architecture of the Deep neural network varying with number of available base stations can be a promising solution to be investigated. Moreover the future work should include studying the proposed prediction scheme performance for more RRM algorithms and in different possible scenarios and cellular cell types.

\section*{Acknowledgment}
This work was supported by project No. LTT18007 funded by Ministry of Education, Youth and Sports, Czech Republic and by the project No. SGS17/184/OHK3/3T/13 funded by CTU in Prague.
The work of David Gesbert was partially supported by HUAWEI-EURECOM Chair on Advanced Mobile Systems towards 6G.

\end{document}